\documentclass[a4paper,10pt,twoside]{cpc-hepnp}

\usepackage{multicol}
\usepackage{graphicx}
\usepackage{booktabs}
\usepackage{amssymb,bm,mathrsfs,bbm,amscd}
\usepackage[tbtags]{amsmath}
\usepackage{lastpage}

\newcommand{\myfig}[4]{
	\begin{center}
	\includegraphics[width=#2 \linewidth]{#1}
	\figcaption{\label{fig:#3} #4} 
	\end{center} }
	
\newcommand{\mygt}[1]{\begin{eqnarray} #1\end{eqnarray}}

\begin{document}

\fancyhead[c]{\small Submitted to 'Chinese Physics C'} \fancyfoot[C]{\small 010201-\thepage}
\footnotetext[0]{Received XXXX}

\title{Bunch evolution study in optimization of MeV ultrafast electron diffraction\thanks{Supported by National Natural Science
Foundation of China (11127507 and 10925523) }}

\author{%
      LU Xian-Hai$^{1,2}$
      \quad DU Ying-Chao$^{1,2}$
      \quad HUANG Wen-Hui$^{1,2}$
      \quad TANG Chuan-Xiang$^{1,2,1)}$\email{Tang.xuh@tsinghua.edu.cn}%
}

\maketitle

\address{%
$^1$ Department of Engineering Physics, Tsinghua University, Beijing 100084, China\\
$^2$ Key Laboratory of Particle  Radiation Imaging (Tsinghua University), Ministry of Education, Beijing 100084, China\\
}

\begin{abstract}
Megaelectronvolt ultrafast electron diffraction (UED) is a promising detection tool for ultrafast processes. The quality of diffraction image is determined by the transverse evolution of the probe bunch. In this paper, we study the contributing terms of the emittance and space charge effects to the bunch evolution in MeV UED scheme, employing a mean-field model with an ellipsoidal distribution as well as particle tracking simulation. Small transverse dimension of the drive laser is found critical to improve the reciprocal resolution, exploiting both smaller emittance and larger transverse bunch size before the solenoid. The degradation of reciprocal spatial resolution caused by the space charge effects should be carefully controlled.
\end{abstract}

\begin{keyword}
ultrafast electron diffraction, reciprocal spatial resolution, space charge effects, bunch evolution
\end{keyword}

\begin{pacs}
07.78.+s, 41.75.-i, 41.85.-p
\end{pacs}

\footnotetext[0]{\hspace*{-3mm}\raisebox{0.3ex}{$\scriptstyle\copyright$}2013
Chinese Physical Society and the Institute of High Energy Physics
of the Chinese Academy of Sciences and the Institute
of Modern Physics of the Chinese Academy of Sciences and IOP Publishing Ltd}%

\begin{multicols}{2}

\section{Introduction}
With the advancement in research on microcosmic and ultrafast processes, a remarkable demand for new tools for direct visualization arises. Discoveries are made using ultrashort X-ray diffraction in the biology and solid-state physics field \cite{schotte-03-science, fritz-07-science}. Compared to large X-ray facilities such as the Linac Coherent Light Source (LCLS) \cite{emma-10-natphot}, ultrafast electron diffraction (UED) is an ideal tool for larger cross-section and tabletop-scale convenience \cite{ihee-01-science, siwick-03-science, ruan-04-science}. In the pump-probe experiment, electron bunch of sub picosecond length can be generated and sent to the interested sample to record its structural information with the diffraction pattern. When the sample is pumped by an intense laser pulse, by adjusting the arrival time of the electron bunch with respect to the pumping time, a series of diffraction patterns are obtained and the dynamical process is retrieved. 

Currently most UED systems employ high-voltage static electric field to accelerate the electron bunch to a range of 30 keV to 60 keV. In this scheme, a major hurdle is the dramatic longitudinal expansion caused by significant space charge effects, and the bunch length puts an up limit of temporal resolution of the system \cite{siwick-05-ol}. To alleviate the impact of space charge effects, the electron flux is limited to 3000--6000 electrons per bunch, which results in a sub-picosecond resolution at the price of hundreds of repeating pumping, required by the signal-to-noise ratio (SNR) for resolving diffraction pattern\cite{siwick-03-science}. One proposed solution is generating a long bunch containing millions of electrons and then compressed the bunch to a sub-picosecond length at the sample \cite{oudheusden-07-jap, tokita-10-prl, chatelain-12-apl}. An alternative method circumvents the space charge effects by accelerating the electrons to megaelectronvolt (MeV) energy using a radio frequency (RF) gun \cite{wang-06-jkps}. In an accelerating field as high as 60 MV/m, the bunch can be boosted to the relativistic regime in several centimeters before significant longitudinal expansion occurs \cite{hastings-06-apl, li-09-nima, musumeci-10-apl,murooka-11-apl}. Benefitting from the development and success of the free electron laser as well as the synchronization system, the RF photocathode gun turns a promising candidate for the UED systems.

The primary concern of the UED system is the ultrashort bunch length, which have been studied by plenary models and simulations \cite{siwick-02-jap, collin-05-jap, reed-06-jap}. These studies focus on the longitudinal evolution of bunch, considering the transverse evolution as a part of model, if not neglecting it at all. In the respect of experiment, the transverse dimension of the probe bunch is of equivalent importance. The spot size on detector determines the the reciprocal spatial resolution of the diffraction patterns (see the following text). With poor reciprocal spatial resolution, crucial details of diffraction pattern will be overwhelmed and lost. In the context of static electric-field-based keV UED scheme, simulation studies on pattern displacement and distortion have been conducted\cite{chatelain-12-ultramicroscopy}. In this paper, we study the transverse evolution of the bunch in the RF gun-based MeV UED scheme with analytical models dedicated to this issue, as well as in simulations method. It should be noted that RF gun has been studied by the accelerator community for decades as an intense electron source for subsequent acceleration, but not yet as an independent diagnostic instrument generating bunch of low charge (few pC) and ultrashort length (sub picosecond).  Here, we present a thorough analysis of factors on reciprocal spatial resolution in the context of the MeV UED system based on Tsinghua S-band MeV UED setup, considering  the RF gun, the focusing solenoid, and the parameters of the drive laser.

The rest of this paper is organized as follows: Section 2 explains the models of the transverse bunch evolution for the MeV UED system, including the underlying principle. Section 3 presents the results of the simulation and corresponding discussion.

\section{Models of transverse evolution}
A schematic of the MeV UED setup is shown in Fig.~\ref{fig:schematic}. The electron bunch is generated by a 1.6 cell S-band RF photocathode gun, which is synchronized with the laser system\cite{yan-09-cpc}. A solenoid coil is attached next to the RF gun for restricting the bunch transversely and for emittance compensation. The solenoid is followed by the sample of interest downstream. A detector camera is implemented at the end of the beamline. The positions of elements are presented in Table~\ref{tab:position}. To accelerate the bunch to a kinetic energy of $E_k$ = 2.8 MeV with a small energy spread, an launching phase of 20 degree and a 60 MV/m peak field at the cathode are chosen. More parameters of operation and probe electrons based on the Tsinghua S-band MeV UED setup can be found in \cite{li-09-cpc}.

\myfig{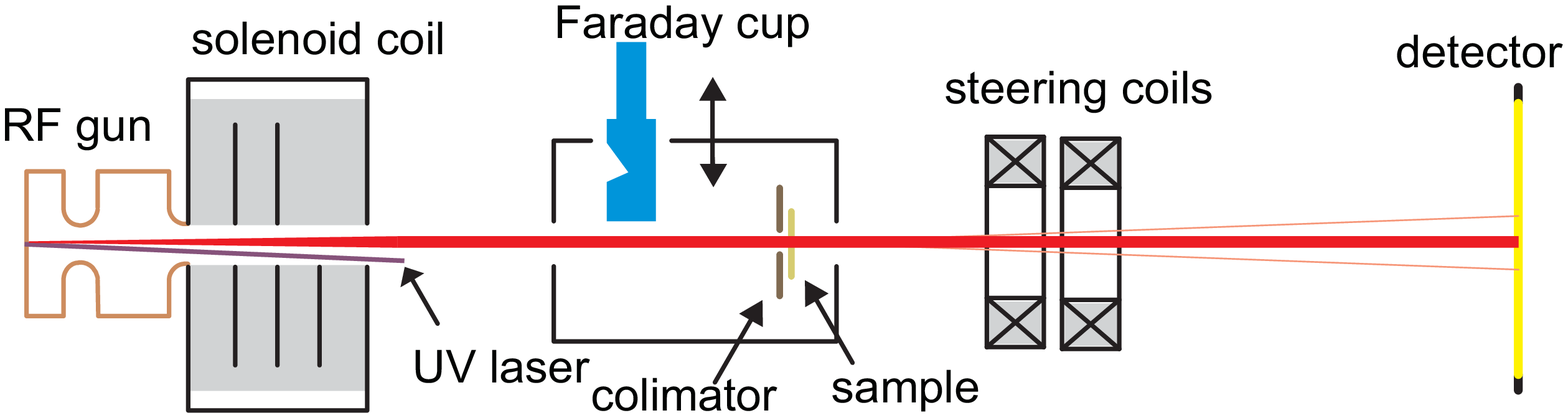}{1}{schematic}{Schematic of the MeV UED setup.}

\begin{center}
\tabcaption{ \label{tab:position}  Positions of elements along the beamline.}
\footnotesize
\begin{tabular*}{80mm}{c@{\extracolsep{\fill}}ccc}
\toprule Elements & Positions   & Units\\
\hline
Cathode & 0 & cm \\
Solenoid & 22 & cm \\
Sample & 73 & cm\\
Detector & 373 & cm\\
\bottomrule
\end{tabular*}
\end{center}

For diffraction patterns, sharper patterns indicate better reciprocal spatial resolution, which can be quantified by the ratio $\mathcal{R}$ of the diameter $R$ and the width $\Delta R$ of the Debye--Scherrer ring \cite{book-grivet-eo}. The ring width $\Delta R$ is determined by the full width at half maximum of the transversal size of the undiffracted beam at the detector, which is 2.35 times of the root mean square (rms) spot size $\sigma_x$ for a Gaussian distribution \cite{musumeci-10-rsi}. Then, the reciprocal spatial resolution $\mathcal{R}$ is:
\mygt{\mathcal{R}=\frac{R}{\Delta R}=\frac{2\theta \alpha d}{2.35\sigma_x}, \label{def-resolving}
}
where $\sigma_x$ is the rms spot size of the bunch at the detector, $\theta$ is the diffraction angle, $d$ is the distance from solenoid to detector, and $\alpha d$ is the distance from sample to detector. According to Eq.~\ref{def-resolving}, the spot size on detector $\sigma_x$ should be minimized to obtain optimal reciprocal resolution, since the system setup ($\alpha d$) and characteristics of the sample ($\theta$) are fixed in the experiment.

\subsection{Model without space charge effects}
First, we consider the situation without space charge effects. Using transfer matrix and thin-lens approximation for the solenoid, we can derive the coordinate of the electron at the detector as:
\mygt{\left[ \begin{array}{c} x \\ x' \end{array}\right]=\left[ \begin{array}{cc} 1 & d \\ 0 & 1 \end{array}\right]
\left[ \begin{array}{cc} 1 & 0 \\ -1/f & 1 \end{array}\right]
\left[ \begin{array}{c} x_0 \\ x_0' \end{array}\right],}
where $d$, $f$, $x$ and $x'$ denote the distance between the solenoid to detector, focal length of the solenoid, as well as position and deviation angle from the axis. The subscript 0 denotes the position before the solenoid. Then the rms value of $x$ can be derived as:
\mygt{\sigma_x=\sqrt{\sigma_{x_0}^2\left( \frac{d}{f}-\frac{\sigma_{x_0}^2+d\sigma_{x_0, x_0'}}{\sigma_{x_0}^2}\right)^2+\frac{\epsilon_x^2d^2}{\sigma_{x_0}^2}}, \label{sigma-x}
}
where $\sigma_x=\sqrt{<x^2>}$, $\sigma_{x x'}=<x x'>$ and the emittance $\epsilon_x=\sqrt{\sigma_{x_0}^2 \sigma_{x'_0}^2-\sigma_{x_0, x_0'}^2}$. The $<>$ defines the second central moment of the particle distribution \cite{floettmann-03-prstab}. The minimal value of Eq.~\ref{sigma-x} can be obtained with proper $f$ as:
\mygt{
\sigma_{x,min}=\frac{\epsilon_x d}{\sigma_{x_0}}. \label{spot-min}
}
Then we substitute Eq.~(\ref{spot-min}) into Eq.~(\ref{def-resolving}) to obtain the expression of spatial resolution without space charge effect:
\mygt{\mathcal{R}_{max}=\frac{2\theta\alpha \sigma_{x_0}}{2.35\epsilon_x}. \label{x-min}
} 

\subsection{Model with space charge effects}
When the space charge effects is significant, the model without Coulomb force underestimates the spot size and the transverse evolution depends on the charge density. In turn, the charge density depends on the dimension of the bunch, resulting in the coupling of the expansion rates in both dimensions. This close loop feature suggests differential equations are proper for modeling the space charge effects. It should be noted that in accelerator community where RF gun is regarded as source followed by acceleration and focusing, attention are usually focused on emittance or peak current, instead of the spot size. Here we perform a dedicated analysis for the issue of spot size optimization.

First, we assume the bunch of a uniform three-dimensional ellipsoidal distribution, which maintains the shape of bunch as well as the uniformity of charge in linear field\cite{luiten-04-prl}. The ellipsoidal bunch can be generated by intense short laser with ``half-circle" radial profile \cite{luiten-04-prl}. The blow-out regime\cite{musumeci-08-prl} is not a widely used regime for conventional photo injector due to limited current, but the feature of uniform ellipsoidal distribution is favorable for UED system. We denote the semi-principal axis $R$ in the transverse direction and the semi-principal axis $L$ in the longitudinal direction. In the rest frame, the transversal electric field in the ellipse is \cite{luiten-04-prl}:
\begin{align}
E_x=& \frac{\rho_0}{\varepsilon_0}M_x x,\label{exelip}\\
M_x=&\frac{1}{2}\left[1-\frac{1+\Gamma^2}{\Gamma^3}\left(\Gamma-\arctan \Gamma\right)\right], \label{el_fa}
\end{align}
where $\rho_0$ is the charge density, $\varepsilon_0$ is the permittivity of vacuum, and $\Gamma=\sqrt{R^2/L^2-1}$ is the eccentricity. We derive the equations of motion for the particles at the transverse and longitudinal edges of the ellipse [with coordinates of ($R$, 0, 0) and (0, 0, $L$)]. Then, we obtain the evolution of the transverse semi-principal axis:
\mygt{
\frac{\mathrm{d}^2 R}{\mathrm{d} t^2}=\frac{4QeM_x}{3 \pi RL \varepsilon_0 m}, \label{rmotion_rf}
}
where $Q$ is the charge of the bunch, $e$ is the charge of electron, and $m$ is the mass of the electron. Then, we transform the coordinate ($x,~y,~z,~t$) from the rest frame of the bunch to the laboratory frame with $R'=R$, $t'=\gamma t$, $L'=L/\gamma$ and $\mathrm{d}^2 z'/\mathrm{d} t'^2=\beta^2 c^2$. The result is:
\mygt{\frac{\mathrm{d}^2 R'}{\mathrm{d} z'^2}=\frac{4Qe^2 M'_x}{3\pi R'L' \varepsilon_0 m\beta^2 \gamma^3 c^2 }, \label{rmotion_lf}
}
where $\gamma$ is the Lorentz factor, $c$ is the velocity of light in vacuum, $\beta$ is the velocity of the bunch relative to the speed of light $c$. Similarly, we obtain the equation of motion for the longitudinal direction:
\mygt{
\frac{\mathrm{d}^2 L'}{\mathrm{d} z'^2}=\frac{4Qe^2M'_z}{3 \pi R'^2 \varepsilon_0 m\beta^2 \gamma^3 c^2 }. \label{L-equation}
}

When excluding the effect of space charge, we use the equation of transverse motion with magnetic field $B_z$  of the solenoid and emittance $\epsilon_x$ \cite{book-lee-particle} as:
\mygt{\frac{\mathrm{d}^2 \sigma_x}{\mathrm{d} z^2}+k_0^2\sigma_x- \frac{\epsilon_{x}^2}{\sigma_x^3}=0 \label{betatron}\\
k_0=\frac{eB_z}{2\gamma m \beta c}.}
For the uniform ellipsoidal charge distribution, the relation between the semi-principal axis and rms value is $\sigma_x=R/\sqrt{5}$. Then, we add the space charge term of Eqs. (\ref{rmotion_lf}) to (\ref{betatron}) and obtain the radial differential equation for the ellipsoidal bunch (we omit ``~'~'' here for the laboratory frame):
\mygt{\frac{\mathrm{d}^2 R}{\mathrm{d} z^2}=k_0^2R-\frac{25\epsilon_{x}^2}{R^3}+\frac{4Qe^2M_x}{3 \pi RL \varepsilon_0 m\beta^2 \gamma^3 c^2 }.\label{rmotion_lf_n}}

Combining Eqs. (\ref{L-equation}) and (\ref{rmotion_lf_n}), we can investigate the evolution of the transverse size of the bunch quantitatively and evaluate the space charge effects under specified charge. The result is discussed in Section 3.

\subsection{Simulation code}

We use the particle tracking code ASTRA \cite{code-astra} for simulation purpose. ASTRA is a particle-in-cell code for photoinjector simulation, where different initial distributions of the bunch on the photocathode can be specified. In the simulation, the space charge effects can be switched off for comparison purpose. Using ASTRA, we can obtain bunch parameters such as $\sigma_x$, $L$, and expansion rates at the exit of the RF gun. These parameters can be used as initial values for the transfer matrix model without space charge effects or the differential model with space charge effects. We can also track the bunch dimensions from start to end to check the validity of the models above.

\section{Results and discussion}

\subsection{Model without space charge effects}
First we generate  bunch distributions of different thermal emittance at the cathode  and switch off the space charge effects in ASTRA. Given the absence of the space charge effects and the weak nonlinear field of the RF gun for a sub-picosecond bunch, the emittance is preserved from the cathode \cite{carlsten-89-nima}. For each specified emittance, we conduct solenoid scanning and obtain the minimized $\sigma_x$. In the transfer matrix model, we choose the middle of the solenoid as the thin-lens plane and use the spot size $\sigma_{x_0}$ from simulation as input values. The results of the model and simulation are compared in Fig.~\ref{fig:no-sc-compare}.
\myfig{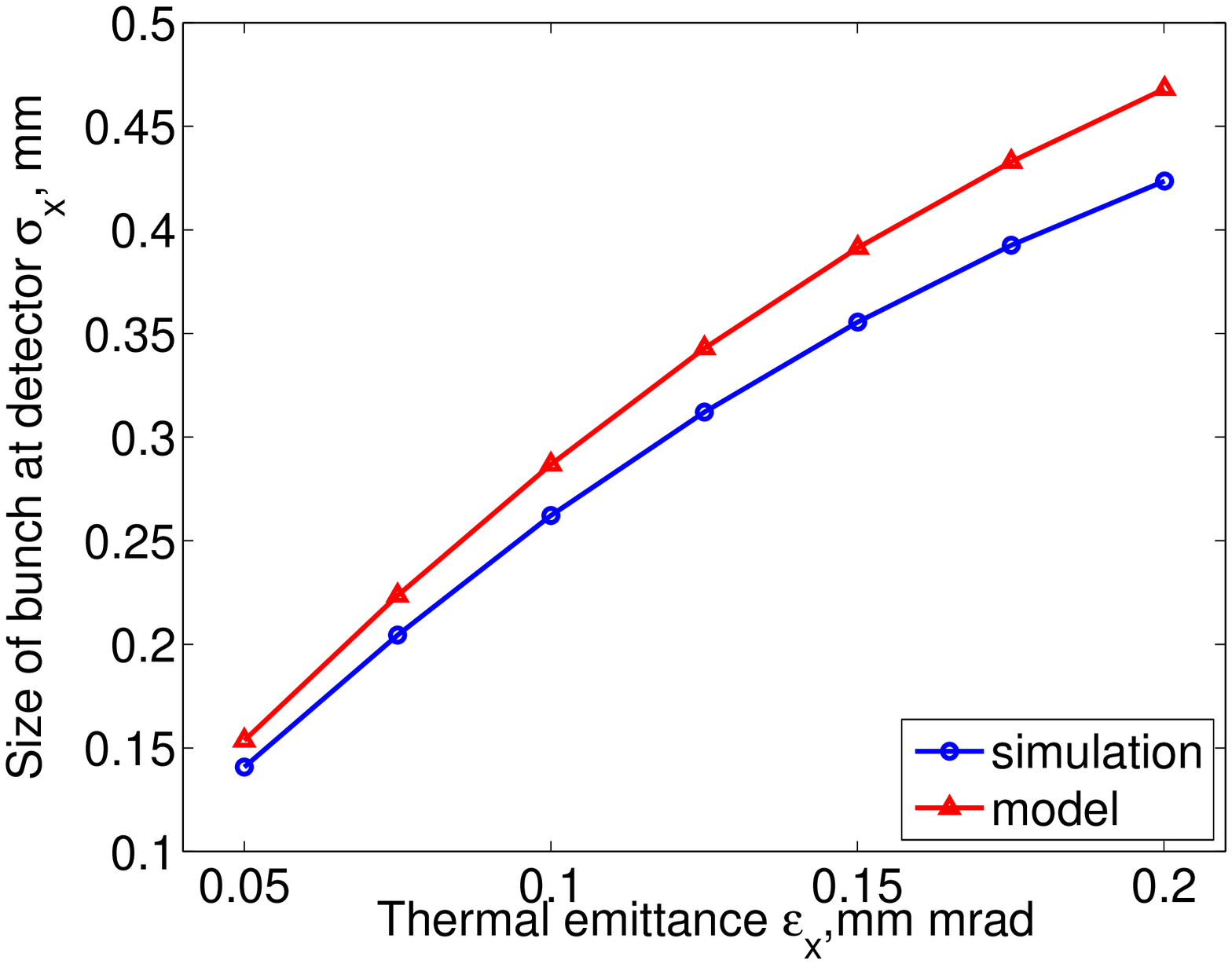}{1}{no-sc-compare}{Optimized transverse size of the bunch at the detector determined by transfer matrix model and ASTRA result without space charge effect.}

The results have similar trends and the discrepancy between both results is within 10\%. As indicated by Eq.~(\ref{spot-min}), the transverse size on the detector $\sigma_x$ decreases with emittance $\epsilon_x$. The small discrepancy between the model and the simulation may be attributed to the effective distance $d$ and the thin-lens approximation. Using this model, we can also estimate the influence of energy spread on $\sigma_x$. Firstly, the focal length of the solenoid can be derived as:
\mygt{f=\frac{1}{\int(eB_z/2\beta \gamma mc)^2 dz}.}
Then the smearing of ring caused by the energy spread can be calculated as:
\mygt{\frac{\Delta x}{x}=-\frac{\Delta f}{f}=-\frac{2\gamma \Delta \gamma}{\gamma^2-1}.}
For a typical relative energy spread ($\Delta \gamma/\gamma$) smaller than 1$\%$ for the MeV UED system, the effect of the energy spread on $\sigma_x$ can be neglected.

\subsection{Model with space charge effects}
The mean-field model has been proposed and proved effective and efficient for  modeling the evolution of the bunch length and energy spread, in keV UED regime \cite{siwick-02-jap, collin-05-jap, reed-06-jap}. In the mean-field model, the bunch is modeled as a cylinder with evolving aspect ratio smaller than 1 (a thin disk) in the region before the sample. These studies focus on the dramatic longitudinal expansion before the sample as the first concern. However, for the MeV UED regime, the longitudinal expansion is mitigated by the relativistic energy thus not as severe as in the keV UED regime. On the other hand, the transverse expansion in the much longer drifting distance (several meters in MeV system compared with several cm in keV system) results significant degradation of reciprocal spatial resolution. The significant influence of transverse evolution should be taken into consideration of model construction.

In the ellipsoidal-distribution-based model introduced above, firstly we choose initial values of ($R$, ${dR}/{dz}$) and ($L$, ${dL}/{dz}$) from the simulation result of ASTRA at $z$ = 0.5 m, corresponding to the exit of the solenoid. In this case, we examine the performance of the model in free space. Furthermore, we select another start point upstream at $z$ = 0.1 m corresponding to the exit of the gun, including the solenoid scanning in the model. The results of both starting points are presented in Fig.~\ref{fig:kv-sole}, compared with ASTRA simulation. The consistency between the model and simulation is excellent for region either with or without the magnetic field.
\myfig{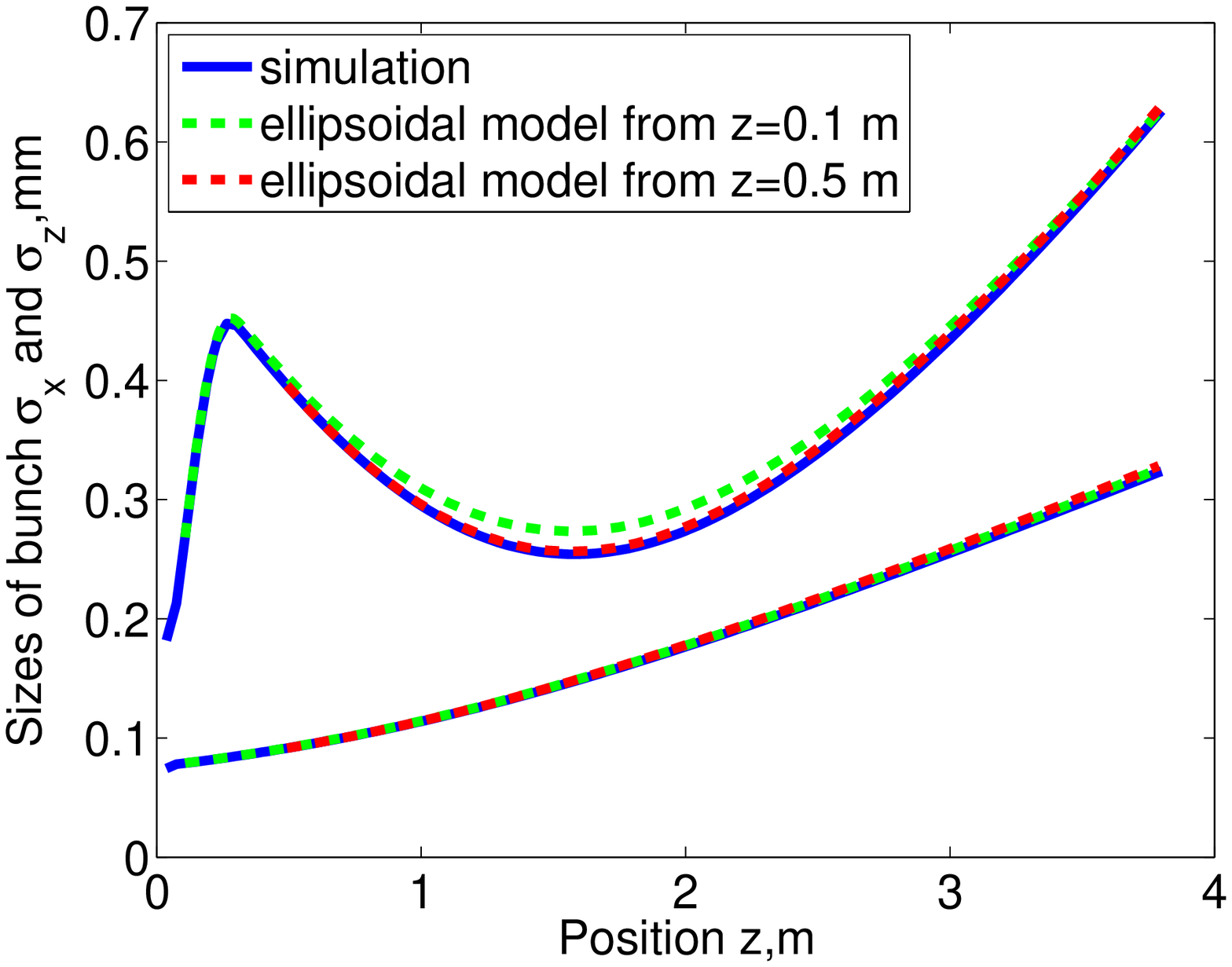}{1}{kv-sole}{Transverse (top) and longitudinal (bottom) evolutions of the bunch for three cases: simulation (starting from z=0 m), model from the exit of RF gun (starting from z=0.1 m), and model from the exit of solenoid (starting from z=0.5 m).}

With the help of differential model, we evaluate the space charge effects by examining the extent of transverse expansion for different charge. We choose the exit of the RF gun as the start point and obtain the minimal spot size under different charges. The result (shown in Fig.~\ref{fig:drift-sole-Q}) indicates that the expansion caused by the space charge effect is almost linear with respect to charge. In the case of $Q$ = 1 pC, the optimized spot size is four times larger than that in the case of $Q$=0 due to space charge effects, under identical initial distribution. While limiting $Q$ is required to improve temporal resolution in keV UED system \cite{siwick-02-jap}, here we draw the conclusion that to improve the performance of MeV UED system, especially in the respect of spatial reciprocal resolution, transverse expansion due to space charge effect must to carefully handled. Otherwise, this issue will hinder the hard-earned advantage of favorable SNR feature of the MeV UED system.
\myfig{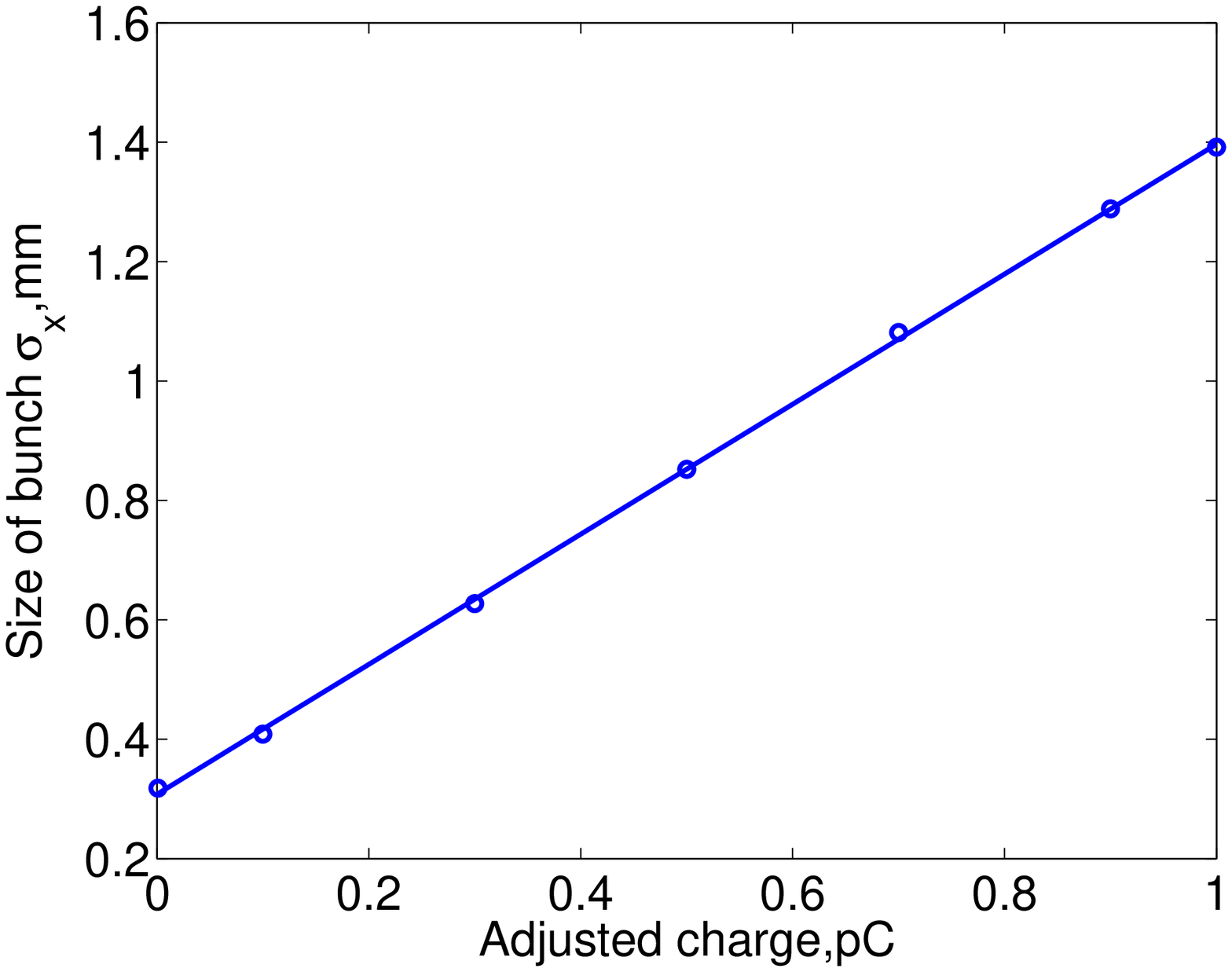}{1}{drift-sole-Q}{Optimized transverse size and the linear fitting of different charges calculated by differential model.}

\subsection{Optimization of laser spot size}
Besides the optimization of drifting region discussed above, the parameters of the driving laser is also critical. One consideration of the probe bunch is the emittance, consisting the thermal emittance and the emittance growth \cite{carlsten-89-nima}. The thermal emittance is proportional to the rms laser spot size $\sigma_{laser}$ on the cathode, and takes up a significant part in the low charge regime like UED system. Meanwhile the intensity of laser influences the charge density in the drifting region, thus the bunch size evolution, which will be discussed in the following part. Since the dynamics due to nonlinearity of RF field is beyond the capability of the models, we study this issue relying on the particle tracking code, from the cathode to the detector. The result of optimized spot size with respect to the laser spot size is shown in Fig.~\ref{fig:spot-laser}. Other parameters of laser are listed in Table~\ref{tab:laser}.

\myfig{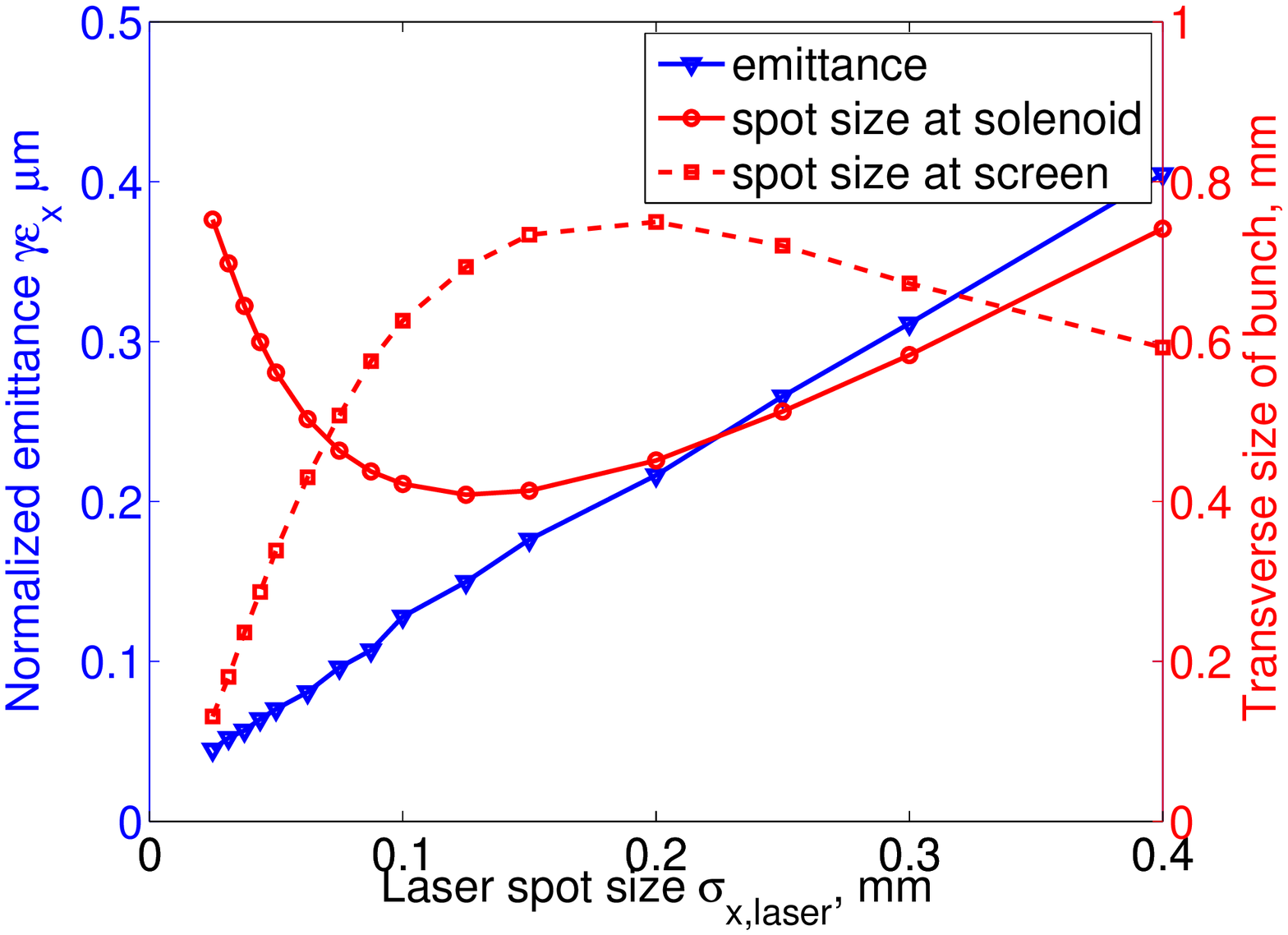}{1}{spot-laser}{Optimized transverse size with respect to spot size of laser on cathode.}
\begin{center}
\tabcaption{ \label{tab:laser} Simulation parameters of the laser on cathode.}
\footnotesize
\begin{tabular*}{80mm}{c@{\extracolsep{\fill}}cc}
\toprule 
Parameter	&	Value \\
	\hline
	Transversal distribution	&	flattop	\\
	Laser spot size $\sigma_{x,laser}$ (rms)		&	0.1 mm	\\
	Longitudinal distribution	&	Gaussian	\\
	Pulse length $\sigma_{z,laser}$ (rms)	&	300 fs	\\
	Beam charge Q		&		1 pC \\
	Thermal emittance $\gamma \epsilon_x$	&	1 $\mu$m/mm \\
\bottomrule
\end{tabular*}
\end{center}

The curve of spot size at screen (red dash line) shows a convex shape with a peak at moderate laser spot size. To illustrate this result, we scrutinize the components of emittance and focusing effect (blue and red solid line respectively in Fig.~\ref{fig:spot-laser}). The extremely small spot size at screen in bottom left corner can be explained by 1) the large transverse size at solenoid caused by high charge density in the gun and 2) the small emittance proportional to laser spot size. This is in consistent with Eq.~\ref{spot-min}, which has emittance $\epsilon_x$ in the numerator and size $\sigma_{x_0}$ in the denominator. When considering space charge effects, the effect is favorable due to lower charge density (indicated by large spot size at solenoid) during drifting. As the laser spot size increases, the emittance term increases monotonously and a turning point of drifting bunch size occurs. The superposition of both effects explains the shape of spot size at screen. Notably, the maximum of spot size at screen is lagged behind the minimal of spot size at solenoid. The lag also confirms the influence of the increasing emittance.

Based on the analysis above, we conduct a thorough simulation for different laser spot size, including the diffraction process of polycrystalline aluminum film using the kinematic method. To optimize the resolution, low  charge $Q$ = 0.1 pC to avoid severe space charge degradation. The result is shown in Fig.~\ref{fig:diff}.
\myfig{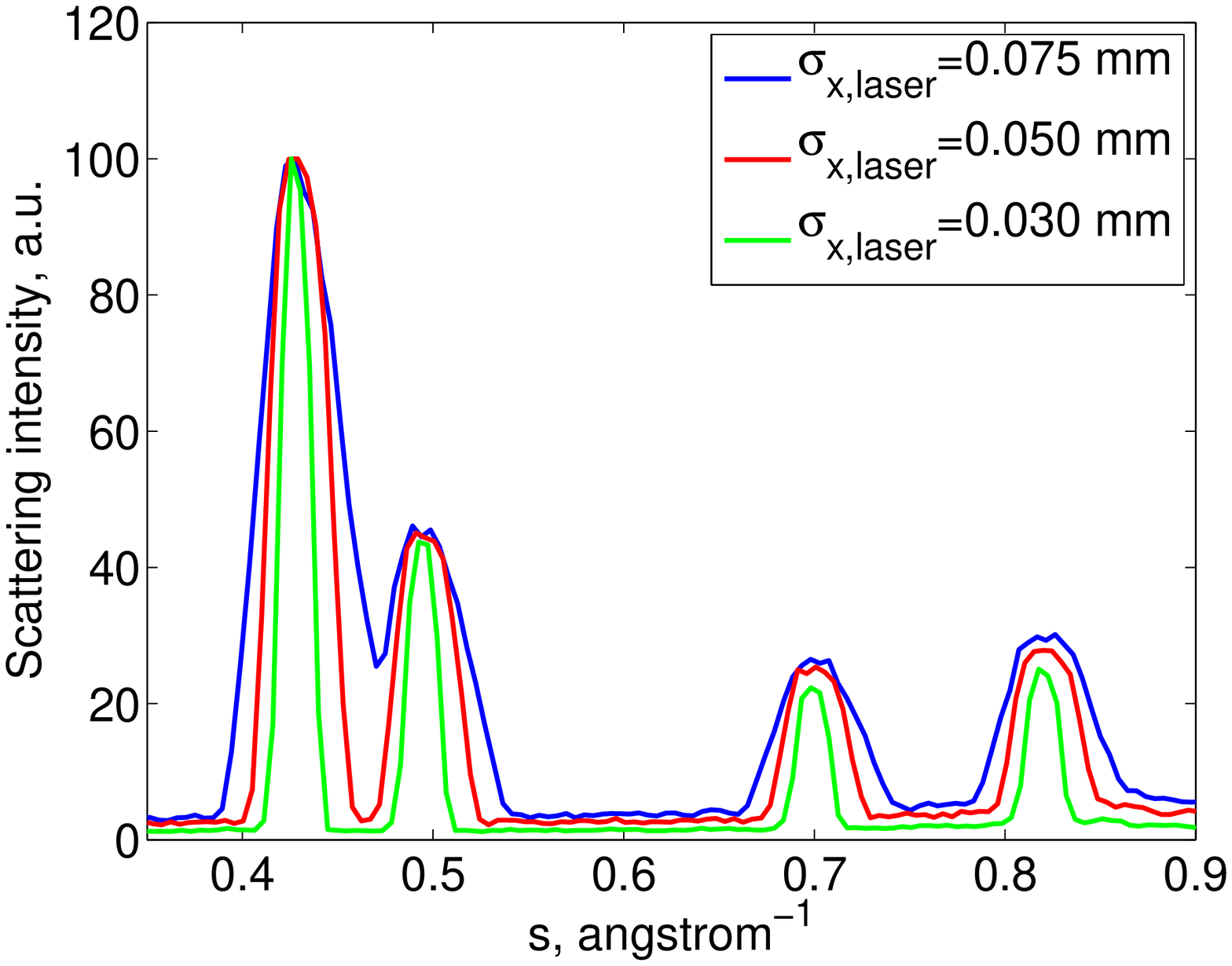}{1}{diff}{Scattering intensity of the diffraction pattern of polycrystalline aluminum with respect to laser spot size. The momentum transfer s is expressed as $s=2\sin(\theta/2)/\lambda$, where $\theta$ is the scattered angle, and $\lambda$ is the de Broglie wavelength of the electron.}

Fig.~\ref{fig:diff} shows that the reciprocal spatial resolution of the diffraction pattern is significantly improved and much sharper than prior simulation results \cite{li-09-rsi} or high quality experiments \cite{musumeci-10-rsi,li-10-ipac}. For the (111) order of the ring, the optimized reciprocal spatial resolution in the simulation is about $\mathcal{R}_{max}$ = 48. This value is three times better than the obtained experimental value of about $\mathcal{R}_{max}$ = 15 for a single shot experimental \cite{li-10-ipac}. Moreover, this value is at the same level as the 20-shot averaged case, which is about $\mathcal{R}_{max}$ = 44 \cite{ murooka-11-apl}. The resolving ability can be further improved to $\mathcal{R}_{max}$ = 78 by using the charge of 1 pC, relaxing the requirement of temporal resolution.

\section{Conclusion}
In this paper, we focused on the transverse evolution of the probe electron bunch in MeV UED system to improve the reciprocal spatial resolution. We introduced transfer matrix to model the bunch evolution without space charge effects, the results of which show that the minimal spot size of the bunch at the detector is proportional to the transversal size of the bunch before the solenoid divided by the emittance. To include the space charge effects, the mean-field model based on the ellipsoidal distribution is modified for characterizing the three dimensional evolutions of the bunch, dedicated for the MeV UED system. This model shows excellent consistency and evaluates the dominating space charge effects in dramatic transverse expansion during the long drifting region. The critical effect of laser spot size on cathode is also examined. The small emittance and the large transverse size at the solenoid, as well as immitigated space charge effects, contribute to the extremely small spot size at screen.

Based on the analysis, we suggest a set of parameters for experiment to significantly improve the reciprocal spatial resolution. Generally, the models are applicable for different MeV UED setups and the analysis offers a guideline for improving the reciprocal spatial resolution of the MeV UED system.

\section{Acknowledgements}
The authors acknowledge the helpful discussions with Dr.~QIAN Hou-Jun, Dr.~LI Ren-Kai and XU Xin-Lu. This work was supported by the National Natural Science Foundation of China (Grant Nos. 11127507 and 10925523).

\vspace{2mm}

\begin{thebibliography}{90}

\bibitem{schotte-03-science}
Schotte F, Lim M, Jackson T A, et al. Science, 2003, {\bf 300}: 1944---1947.

\bibitem{fritz-07-science}
Fritz D M, Reis D A, Adams B, et al. Science, 2007, {\bf 315}: 633---636.

\bibitem{emma-10-natphot}
Emma P, Akre R, Arthur J, et al. Nat. Photonics, 2010, {\bf 4}: 641---647.

\bibitem{ihee-01-science}
Ihee H, Lobastov V A, Gomez U M, et al. Science, 2001, {\bf 291}: 458---462.

\bibitem{siwick-03-science}
Siwick B J, Dwyer J R, Jordan R E, et al. Science, 2003, {\bf 302}: 1382---1385.

\bibitem{ruan-04-science}
Ruan C Y, Lobastov V A, Vigliotti F, et al. Science, 2004, {\bf 304}: 80---84.

\bibitem{siwick-05-ol}
Siwick B J, Green A A, Hebeisen C T, et al. Opt. Lett., 2005, {\bf 30}: 1057--1059.

\bibitem{oudheusden-07-jap}
Oudheusden T V, Jong E D, Geer S V, et al. J. Appl. Phys., 2007, {\bf 102}: 093501.

\bibitem{tokita-10-prl}
Tokita S, Hashida M, Inoue S, et al. Phys. Rev. Lett., 2010, {\bf 105}: 215004.

\bibitem{chatelain-12-apl}
Chatelain R P, Morrison V R, Godbout C, et al. Appl. Phys. Lett., 2012, {\bf 101}: 081901.

\bibitem{wang-06-jkps}
WANG Xuan, XIANG Dao, Kim T, et al. J. Korean Phys. Soc. 2006, {\bf 48}: 390.

\bibitem{hastings-06-apl}
Hastings J B, Rudakov F M, Dowell D H, et al. Appl. Phys. Lett., 2006, {\bf 89}: 184109.

\bibitem{li-09-nima}
LI Ren-Kai, TANG Chuan-Xiang. Nucl. Instr. and Meth. A, 2009,
  {\bf 605}: 243---248.

\bibitem{musumeci-10-apl}
Musumeci P, Moody J T, Scoby C M, et al. Appl. Phys. Lett., 2010, {\bf 97}: 063502.

\bibitem{murooka-11-apl}
Murooka Y, Naruse N, Sakakihara S, et al. Appl. Phys. Lett., 2011, {\bf 98}: 251903.

\bibitem{siwick-02-jap}
Siwick B J, Dwyer J R, Jordan R E, et al. J. Appl. Phys., 2002, {\bf 92}: 1643--1648.

\bibitem{collin-05-jap}
Collin S, Merano M, Gatri M, et al. J. Appl. Phys., 2005, {\bf 98}: 094910.

\bibitem{reed-06-jap}
Reed B W. J. Appl. Phys., 2006, {\bf 100}: 034916.

\bibitem{chatelain-12-ultramicroscopy}
Chatelain R P, Morrison V, Godbout C, et al. Ultramicroscopy, 2012, {\bf 116}: 86.

\bibitem{yan-09-cpc}
YAN Li-Xin, Du Ying-Chao, DU Qiang, et al. Chin. Phys. C, 2009, {\bf 33 Suppl. II}: 154.

\bibitem{li-09-cpc}
LI Ren-Kai, TANG Chuan-Xiang, HUANG Wen-Hui, et al. Chin. Phys. C, 2009, {\bf 33 Suppl. II}: 165.

\bibitem{book-grivet-eo}
Pierre Grivet. Electron optics. Pergamon Press, 1965.

\bibitem{musumeci-10-rsi}
Musumeci P, Moody J T, Scoby C M, et al. Rev. Sci. Instrum, 2010, {\bf 81}: 013306.

\bibitem{floettmann-03-prstab}
Floettmann K. Phys. Rev. ST Accel. Beams, 2003, {\bf 6}: 34202.

\bibitem{luiten-04-prl}
Luiten O J, Geer S B, Loos M D, et al. Phys. Rev. Lett., 2004, {\bf 93}: 94802.

\bibitem{musumeci-08-prl}
Musumeci P, Moody J T, England R J, et al. Phys. Rev. Lett., 2008, {\bf 100}: 244801.

\bibitem{book-lee-particle}
Lee S Y. Accelerator Physics. World Scientific, 2004.

\bibitem{code-astra}
Floettmann K.http://www.desy.de/~mpyflo/Astra, 2012.

\bibitem{carlsten-89-nima}
Carlsten B E. Nucl. Instr. and Meth. A, 1989, {\bf  285}: 313---319.

\bibitem{li-09-rsi}
LI Ren-Kai, TANG Chuan-Xiang, Du Ying-Chao, et al. Rev. Sci. Instrum, 2009,  {\bf 80}: 083303.

\bibitem{li-10-ipac}
LI Ren-Kai, HUANG Wen-Hui, DU Ying-Chao, et al. Proc. of IPAC'10, 2010, 229.


\end{thebibliography}

\vspace{3mm}

\end{multicols}

\clearpage

\end{document}